\documentclass{kapproc} 

\usepackage{procps} 

\usepackage[dvips]{graphicx}

\upperandlowercase

\kluwerbib 

\begin{document}



\articletitle[Galactic Rotation Using Star Clusters]
{Determination of the Galactic Rotation \\Using Open Star Clusters: Preliminary Results}












\author{Peter M. Frinchaboy \\ and \\ Steven R. Majewski}
\affil{University of Virginia, Department of Astronomy \\
       P.O. Box 3818, Charlottesville, VA 22903-0818}
\email{pmf8b, srm4n@vriginia.edu}









\begin{abstract}
The dark matter distribution of the Milky Way remains among the major
unsolved problems about our home galaxy.  The masses of other spiral
galaxies can be determined from their rotation curves through long-slit
spectroscopy.  But for the Milky Way obtaining the complete rotation
curve is a more complex problem. By measuring the 3-dimensional motions of
tracer objects the rotation curve and
Galactic mass distribution can be derived, even outside the solar circle 
where H$_{\rm I}$ tangent point analysis is not possible.  We present the first
findings from a project to measure the motions of open clusters, 
both inside and outside the solar circle. 
From a nearly uniform sample of spectroscopic data for large numbers of stars in
over 50 open clusters in the third and fourth Galactic quadrants, 
we derive the speed of Galactic rotation at the solar circle as 
$\Theta_0 = 214^{+6}_{-9}$ km s$^{-1}$.
Future work will include clusters in the other Galactic quadrants and 
analysis of the local rotation curve. 
\end{abstract}

 \begin{keywords}Galaxy: open clusters and associations, Galaxy: Fundamental
parameters, Galaxy: Structure, Galaxy: Dynamics 
 \end{keywords}


Considering the importance of galactic potentials in shaping
the chemodynamical evolution of their stellar populations, 
it is unfortunate that the surface mass distribution of the 
Milky Way is still poorly constrained, even while our knowledge of Galactic
stellar abundance distributions grows ever more detailed. 
One reason for this shortcoming is the difficulty of determining 
the Galactic rotation curve in the outer disk.  
The inner Galactic rotation curve has been derived from
tangent point analysis of H$_{\rm I}$ radial velocities (RVs; e.g., Malhotra 1995).  
However, outside the solar circle one must also obtain proper motions {\it and}
reliable distances for some suitable stellar tracer population to properly derive the rotation curve.  
To this end, we have collected a near uniform sample of spectroscopic data from 
stars in over 100 open clusters at medium--high resolution ($R \sim 15,000$) using the 
HYDRA multi-fiber spectrographs on the CTIO 4-meter and WIYN 3.5-meter telescopes.
We specifically target stars for which proper motions  have previously been measured 
(e.g., Dias et al.~2001), and
place additional fibers on other stars in the field
to expand the membership census in anticipation of future astrometric
surveys (e.g., GAIA).    
The average motion of all verified members will yield a reliable, 
accurate systemic proper motion and RV for each cluster to trace the rotation curve.
However, in this proceedings we focus specifically on an analysis of the speed of Galactic rotation 
at the solar circle ($\Theta_0$) based solely on cluster RVs and positions.

\begin{figure}[h]
\includegraphics[width=4.0in]{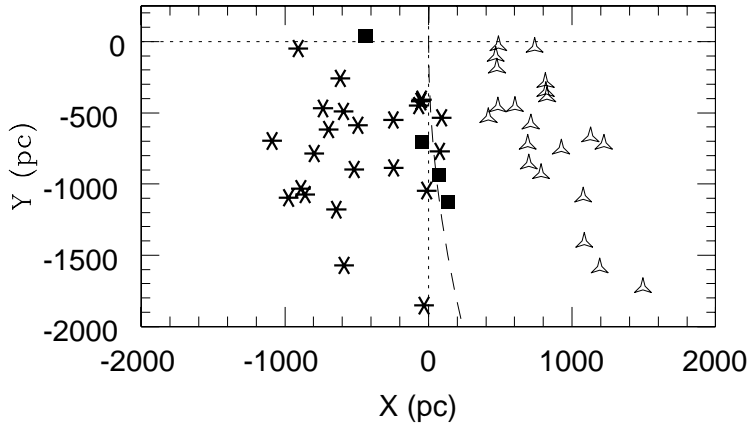}
\caption{Visualization of the Milky Way in the $X,Y$ coordinate system of the Galaxy with the Sun at (0,0). 
The 56 clusters used here (primarily in the third and fourth Galactic quadrants) are shown by their 
$V_{lsr}$:  stars ($V_{lsr} > 5$ km s$^{-1}$), squares ($-5 < V_{lsr} < 5$ km s$^{-1}$), 
and triangles ($V_{lsr} < -5$ km s$^{-1}$).  
The transition between negative and positive $V_{lsr}$ clusters lies near the solar circle (dashed line)
as expected for well-behaved circular Galactic rotation. 
}
\end{figure}
Data for the sub-sample of 56 clusters presented here were all 
obtained using the 
Blanco 4-meter telescope at CTIO. The spectra were reduced using standard techniques. 
RVs were determined using the IRAF fxcor package as described in \cite{pmf_pmf05}.
Cluster star members were isolated both from our RVs and the proper motion 
membership criteria of Dias et al.~(2001, 2002a). 
We adopt a $\pm 5$--10 km s$^{-1}$ RV range to discriminate cluster members, 
which is several times larger than the typical individual stellar 
RV errors (1--2 km s$^{-1}$) and the expected velocity dispersions of these systems. 
With proper motions and RVs for each star we can, in principle, derive full space velocities, 
but we already have significant sensitivity to 
the speed of the local standard of rest through the cluster RVs and positions, for a  
conservative first analysis.  An average of 8--9 stellar 
members per cluster yields a mean error in systemic velocity for all clusters used here of 3.9 km s$^{-1}$. 
The analysis was completed using distances from  
the \cite{pmf_dias02b} catalog combined with our heliocentric RVs 
converted to $V_{lsr}$ (see Figure 1) using the 
solar peculiar motion ($U, V, W$) velocities (10, 5.2, 7.2) km s$^{-1}$ given in \cite{pmf_bm98}.

\begin{figure}[h]
\includegraphics[width=4.3in]{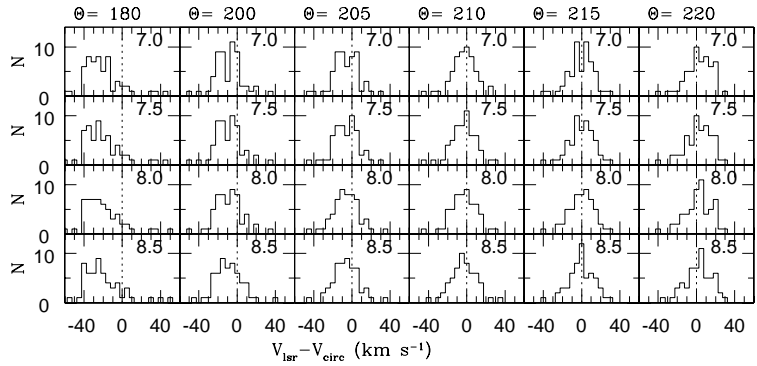}
\caption{Histograms of the difference of $V_{lsr}-V_{circ}$ for the 56 open clusters
for a range of adopted $R_0$ and $\Theta_0$. 
The dotted line denotes a difference of zero which corresponds to a flat rotation curve at the given $\Theta_0$
}
\end{figure}

Under the assumption that open clusters closely trace the circular 
speed of the Milky Way at their Galactic radius, 
we can predict their expected observed $V_{lsr}$ under an assumed 
rotation curve and solar Galactocentric distance ($R_{0}$).
For simplicity, we assume a flat rotation curve over the relevant Galactic radii. 
Comparison of our measured $V_{lsr}$'s to derived $V_{circ}$'s 
(i.e., the $V_{lsr}$ a cluster would have in the  
same position for a flat rotation curve of speed $\Theta_0$) over a range of $R_{0}$ (7.0--8.5 kpc) and  
$\Theta_0$ (180--220 km s$^{-1}$) reveals a systematic trend with assumed $\Theta_0$ but little apparent 
sensitivity to $R_{0}$ (see Figure 2). Over this range of $R_{0}$ the mean 
($V_{lsr}$-$V_{circ}$) difference was consistently closest to zero 
when $\Theta_0$ ranged between only 210--215 km s$^{-1}$. 
With finer analysis of the $\Theta_0=210$--215 km s$^{-1}$ region (see Figure 3), we find that 
($V_{lsr}$-$V_{circ}) = 0.0 \pm 0.2$ km s$^{-1}$ for $\Theta_0 = 214^{+6}_{-9}$ km s$^{-1}$. 
The quoted errors are the 3$\sigma$ deviation given by the error on the mean.   This $\Theta_0$ 
result is in the middle of the wide range of 
other recent findings that range from $\Theta_0=184$ km s$^{-1}$ 
(\cite{pmf_om98}) to $\Theta_0=255$ km s$^{-1}$ 
(\cite{pmf_uohikh00}), two limiting values strongly ruled out by our open cluster velocities 
if the Milky Way has a well-behaved, simple, 
circular velocity field. Future work on this project will nearly double the number of clusters 
available (to approximately 100) and expand coverage to the first and second Galactic 
quadrants. Additionally, the proper motion data will be incorporated to investigate not only 
the mean Galactic rotation {\it curve} but whether second order 
variations in the local velocity field of the disk can be discerned.  
\begin{figure}[h]
\includegraphics[width=4.4in]{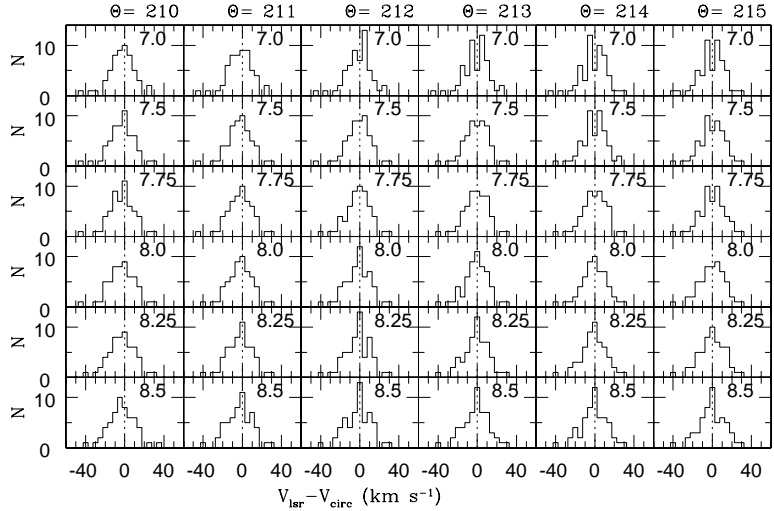}
\caption{
Same as Figure 2 with a finer grid of assumed $\Theta_0$.  We find 
that $\Theta_0=214$ km s$^{-1}$ provides the best match to the Local 
Standard of Rest velocity for our preliminary cluster sample.}
\end{figure}



\begin{acknowledgments}
We would like to thank the NOAO TAC for granting this project long-term status and 
NOAO for travel support of this thesis. 
We acknowledge funding by NSF grant AST-0307851, NASA/JPL contract
1228235, the David and Lucile Packard Foundation, and the F.H. Levinson 
Fund of the Peninsula Community Foundation.  PMF also acknowledges support from 
the Virginia Space Grant Consortium, an AAS International Travel Grant, and the conference organizers.
\end{acknowledgments}



%




\end{document}